\documentclass[conference]{IEEEtran}

\IEEEoverridecommandlockouts

\usepackage[english]{babel} 
\usepackage{microtype}      
\usepackage{csquotes}       
\usepackage[normalem]{ulem} 
\usepackage{xspace}         
\usepackage{textcomp}       

\usepackage{type1cm}        
\usepackage{bold-extra}     
\usepackage{bm}             
\usepackage{bbm}            
\usepackage{bbold}          
\usepackage{siunitx}        
\usepackage{xfrac}          

\usepackage{amsmath,amsfonts,amssymb,amsthm,mathtools} 
\usepackage{physics}        
\usepackage{thmtools,thm-restate} 

\usepackage{booktabs}       
\usepackage{multirow}       
\usepackage{array}          

\usepackage{graphicx}       
\usepackage[style=base, tableposition=top]{caption} 
\usepackage[caption=false,font=normalsize,labelfont=sf,textfont=sf]{subfig} 
\usepackage{tikz}           
\usetikzlibrary{patterns, positioning, arrows, bayesnet, calc} 
\usepackage{tikz-cd}        
\usepackage{pgfplots}       
\usepgfplotslibrary{patchplots}
\pgfplotsset{compat=1.15}   
\usepackage{tcolorbox}      

\usepackage{algorithmic}

\usepackage[bookmarks,colorlinks=true,pagebackref=false,backref=false]{hyperref} 
\usepackage[capitalize]{cleveref} 
\usepackage{cite}

\usepackage{balance}        
\usepackage{hyphenat}       
\usepackage[show]{chato-notes} 
\usepackage{stfloats}       
\usepackage{verbatim}       

{\par\egroup\vskip 0.25ex}

\usepackage[dvipsnames]{xcolor}
\hypersetup{
   colorlinks=true,
   linkcolor={red!50!black},
   filecolor={green!50!black},
   citecolor={green!50!black}, 
   urlcolor={blue!80!black},
}

\definecolor{mypurple}{RGB}{254, 68, 218}
\definecolor{myred}{HTML}{E13D66}
\definecolor{mycyan}{HTML}{70D7D0}
\definecolor{mylightblue}{HTML}{2274A5}
\definecolor{mydarkblue}{HTML}{0C0A3E}

\theoremstyle{plain}

\theoremstyle{definition}

\theoremstyle{remark}

\crefname{theorem}{Thm.}{Thms.}
\crefname{proposition}{Prop.}{Props.}
\crefname{lemma}{lem.}{lems.}
\crefname{corollary}{Cor.}{Cors.}
\crefname{definition}{Def.}{Defs.}
\crefname{section}{Sec.}{Secs.}
\crefname{figure}{Fig.}{Figs.}
\crefname{problem}{Prob.}{Probs.}
\crefname{appendix}{App.}{Apps.}
\crefname{equation}{Eq.}{Eqs.}
\crefname{table}{Tab.}{Tabs.}

\DeclareMathOperator*{\argmin}{arg\,min}

\newcommand{\at}[2][]{#1|_{#2}}
\newcommand{\frob}[1]{\ensuremath{\norm{#1}_{\mathrm{F}}}\xspace}
\newcommand{\prox}{\ensuremath{\mathrm{prox}}\xspace}

\newcommand{\reall}{\ensuremath{\mathbb{R}}\xspace}

\newcommand{\oblique}[2]{\ensuremath{\mathrm{OB}({#1},{#2})}\xspace}
\newcommand{\sphere}[1]{\ensuremath{\mathrm{S}^{{#1}}}\xspace}
\newcommand{\orthogonal}[1]{\ensuremath{\mathrm{O}({#1})}\xspace}

\newcommand{\edgeset}{\ensuremath{\mathcal{E}}\xspace}

\newcommand{\mya}{\ensuremath{\mathbf{a}}\xspace}
\newcommand{\zeros}{\ensuremath{\boldsymbol{0}}\xspace}

\newcommand{\A}{\ensuremath{\mathbf{A}}\xspace}
\newcommand{\B}{\ensuremath{\mathbf{B}}\xspace}

\newcommand{\D}{\ensuremath{\mathbf{D}}\xspace}

\newcommand{\identity}{\ensuremath{\mathbf{I}}\xspace}

\newcommand{\M}{\ensuremath{\mathbf{M}}\xspace}
\newcommand{\myO}{\ensuremath{\mathbf{O}}\xspace}
\newcommand{\myP}{\ensuremath{\mathbf{P}}\xspace}

\newcommand{\R}{\ensuremath{\mathbf{R}}\xspace}
\newcommand{\eS}{\ensuremath{\mathbf{S}}\xspace}

\newcommand{\U}{\ensuremath{\mathbf{U}}\xspace}

\newcommand{\X}{\ensuremath{\mathbf{X}}\xspace}
\newcommand{\Y}{\ensuremath{\mathbf{Y}}\xspace}
\newcommand{\Z}{\ensuremath{\mathbf{Z}}\xspace}

\usepackage{graphicx}

\title{Learning Network Sheaves \\for AI-native Semantic Communication\\
\thanks{The work was supported by the SNS JU project 6G-GOALS under the EU’s Horizon program Grant Agreement No 101139232, and by Huawei Technology France SASU under Grant N. Tg20250616041.}
}
 
 \author{
 \IEEEauthorblockN{
 Enrico Grimaldi\IEEEauthorrefmark{1}\IEEEauthorrefmark{2}, 
 Mario Edoardo Pandolfo\IEEEauthorrefmark{1}\IEEEauthorrefmark{2}, 
 Gabriele D'Acunto\IEEEauthorrefmark{2}\IEEEauthorrefmark{3},
 Sergio Barbarossa\IEEEauthorrefmark{3} and 
 Paolo Di Lorenzo\IEEEauthorrefmark{2}\IEEEauthorrefmark{3}
 }
 
  \IEEEauthorblockA{
  \IEEEauthorrefmark{1}Department of Computer, Control and Management Engineering, Sapienza University of Rome, Italy}
 \IEEEauthorblockA{\IEEEauthorrefmark{2}National Inter-University Consortium for Telecommunications (CNIT), Parma, Italy}
 \IEEEauthorblockA{\IEEEauthorrefmark{3} Department of Information Engineering, Electronics, and Telecommunications, Sapienza University, Rome, Italy}

\IEEEauthorblockA{
\{
enrico.grimaldi,
marioedoardo.pandolfo,
gabriele.dacunto,
paolo.dilorenzo,
sergio.barbarossa\}@uniroma1.it
}
\vspace{-2em}
}

\begin{document}

\maketitle

\begin{abstract}
Recent advances in AI call for a paradigm shift from bit-centric communication to goal- and semantics-oriented architectures, paving the way for AI-native 6G networks. In this context, we address a key open challenge: enabling heterogeneous AI agents to exchange compressed latent-space representations while mitigating semantic noise and preserving task-relevant meaning. We cast this challenge as learning both the communication topology and the alignment maps that govern information exchange among agents, yielding a learned network sheaf equipped with orthogonal maps. This learning process is further supported by a semantic denoising and compression module that constructs a shared global semantic space and derives sparse, structured representations of each agent’s latent space. This corresponds to a nonconvex dictionary learning problem solved iteratively with closed-form updates.
Experiments with multiple AI agents pre-trained on real image data show that the semantic denoising and compression facilitates AI agents alignment and the extraction of semantic clusters, while preserving high accuracy in downstream task. The resulting communication network provides new insights about semantic heterogeneity across agents, highlighting the interpretability of our methodology.
\end{abstract}

\section{Introduction}\label{sec:introduction}

Semantic communication (SC) is emerging as a fundamental enabler of AI-native 6G networks \cite{strinati2024goal}. By transmitting task-relevant semantic representations rather than raw data \cite{kountouris2021semantics}, SC enables ultra-low-latency information exchange, improved efficiency under resource constraints, and inherently goal-oriented protocol design \cite{di2023goal}. At the core of SC lies the semantic representation of data: abstract embeddings that discard low-level signal details  while retaining the information required for downstream tasks. Most existing works adopt the latent representations of deep neural networks (DNNs) as semantic embeddings, as they naturally capture task-relevant structure \cite{xie2021deep}. However, embeddings produced by different DNNs are often misaligned due to model-induced variability—such as differences in initialization, architecture, training data, or tasks—and, in many practical settings, the agents cannot be jointly trained to avoid this mismatch. As a result, these independently learned representations introduce semantic noise, which compromises mutual intelligibility between agents in an SC system \cite{sana2023semantic}. In this context, a fundamental problem is to learn, from heterogeneous embeddings produced by pre-trained AI agents, a communication topology that enables effective and parsimonious information exchange between semantically compatible agents while mitigating semantic noise.

\noindent\textbf{Related works.} Several alignment methods have been developed to mitigate semantic noise. Zero-shot approaches rely on anchors to construct isometry-invariant representations \cite{moschella2022relative}, and have been extended to dynamic and multi-agent contexts \cite{fiorellino2025frame}. Supervised methods learn linear or structured alignment maps \cite{merullo2022linearly, moayeri2023text, maiorca2023latent, lahner2024direct}, using orthogonal Procrustes \cite{wang2008manifold}, ADMM optimization \cite{pandolfo2025latent, di2025federated}, or kernel-based criteria \cite{lample2018word, moayeri2023text}. The functional maps framework has also been generalized for latent space alignment \cite{fumero2024latent}. Other strategies explore optimal transport–based alignment \cite{alvarez2019towards} and neural contrastive map learning \cite{jha2025harnessing}. All these techniques can be integrated with SC systems. For example, they can extend Joint Source–Channel Coding (JSCC) and its deep variant (DJSCC), allowing the management of heterogeneous latent spaces in source and channel encoders/decoders \cite{pannacci2025semantic}. However, most existing works on semantic alignment focus on the pairwise setting, and only a few recent studies extend these ideas to multi-agent scenarios \cite{di2025federated, fiorellino2025frame}. Moreover, none of these approaches address how the communication topology should be learned to support the alignment of heterogeneous embeddings across AI agents--a crucial aspect in practical, large-scale SC systems. 

\noindent\textbf{Contributions.} In this paper, to fill the aforementioned gap, we model the multi-agent communication network as a \emph{network sheaf} valued in the category of vector spaces \cite{curry2014sheaves} with orthogonal linear maps on edges, i.e., an unweighted connection graph. Specifically, we assign \emph{(i)} heterogeneous but equal-dimensional embeddings from AI agents to nodes, and \emph{(ii)} orthogonal alignment matrices to edges. We then introduce a two-stage method to learn both the topology and the associated edge maps while mitigating semantic noise.  
First, we perform \emph{semantic denoising} by solving a dictionary learning problem with sparsity and atom-collinearity penalties, yielding \emph{(i)} a global dictionary that serves as a shared semantic space and \emph{(ii)} the embeddigns representations over the latter. Second, using the embedding representations, we \emph{learn the network sheaf}--i.e., both the communication graph and the orthogonal maps--extending the approach from \cite{di2024learning}. We validate the framework through experiments involving multiple AI agents exchanging latent embeddings derived from real image data. The results demonstrate that the global dictionary enables substantial semantic compression with minimal degradation in downstream task accuracy, while simultaneously revealing intrinsic semantic clusters that the sheaf-learning procedure captures by inferring interpretable and goal-oriented communication topologies. 

\noindent\textbf{Notation.} The set of integers from $1$ to $n$ is $[n]$.  
The matrix of zeros of size $m \times n$ is $\zeros_{m\times n}$, and the identity matrix of size $n\times n$ is $\identity_n$.  
Given a matrix $\A \in \reall^{m \times n}$, its $i$-th column is $[\A]_i$.  
The Frobenius norm of $\A$ is $\frob{\A}$, and the $(p,0)$-norm is the number of nonzero columns of $\A$.  
The manifold of orthogonal matrices of size $d\times d$ is denoted by $\orthogonal{d}$. The oblique manifold, i.e., the product of $n$ spheres $\sphere{d}$ in $\reall^{d+1}$, is denoted by $\oblique{d}{n}$.

\section{Methodology}\label{sec:methods}
This section presents the proposed methodology for learning a communication network under semantic noise, while enabling information compression.  
We first introduce the sheaf-theoretic formulation of multi-agent semantic alignment and topology learning in \Cref{subsec:sheaf_learning}, and then describe the semantic denoising step based on dictionary learning in \Cref{subsec:semantic_denoising}.  

\subsection{Sheaf Learning for Network Semantic Alignment}
\label{subsec:sheaf_learning}

Our multi-agent semantic alignment formulation builds on the sheaf-learning framework introduced in~\cite{di2024learning} and constitutes the first sheaf-theoretic model specifically adapted to SCs.

Consider $V$ neural agents trained on a common dataset of $n$ examples represented by $\Y \in \reall^{f \times n}$.  
Each agent $i \in [V]$ extracts $d$-dimensional embeddings $\X_i \in \reall^{d \times n}$.  
Since the dataset is shared, the columns of $\X_i$ are matched sample-wise across agents, while they differ in the feature space and can be aligned through rigid transformations \cite{moschella2022relative}.  
We model the communication network as a graph $\mathcal{G}=(\mathcal{V},\mathcal{E})$ with $|\mathcal{V}|=V$ and equip it with a sheaf valued in real vecotr spaces, which specifies how latent representations propagate between agents.
\noindent\textit{\textbf{Definition 1 (Network sheaf).}}  
Let $(\Pi,\trianglelefteq)$ denote the face--incidence poset of a graph $\mathcal{G}=(\mathcal{V},\mathcal{E})$.  
A \emph{network sheaf valued in $\mathrm{Vect}_\reall$} is a functor $\mathcal{F} : (\Pi,\trianglelefteq) \rightarrow \mathrm{Vect}_\reall \,$,
assigning to each node $v\in\mathcal{V}$ and edge $e\in\mathcal{E}$ a real vector space $\mathcal{F}(v)$ and $\mathcal{F}(e)$, and to each incidence relation $v\trianglelefteq e$ a linear \emph{restriction map} $\mathcal{F}_{v\trianglelefteq e}:\mathcal{F}(v)\rightarrow \mathcal{F}(e)$.

In our setting, node stalks $\mathcal{F}(v)$ are the spaces where latent representations $\X_v$ live, while edge stalks $\mathcal{F}(e)$ represent abstract communication spaces where semantic information is exchanged.  
The space of $0$-cochains is $C^{0}(\mathcal{G},\mathcal{F})=\bigoplus_{v\in\mathcal{V}}\mathcal{F}(v)\,$,
where the concatenation of all latent representations $\X \coloneqq [\X_1, \ldots, \X_V]$ resides.  
Global consistency is captured by the space of global sections $H^{0}(\mathcal{G},\mathcal{F}) = \ker \delta \,$,
where $\delta$ 
is the coboundary operator acting blockwise as
\[
\delta(x)_e=\mathcal{F}_{u\trianglelefteq e}(x_u)-\mathcal{F}_{v\trianglelefteq e}(x_v)\,,
\qquad \forall \, e=(u,v)\in\mathcal{E}\,.
\]
Analogously to the graph Laplacian, one defines the sheaf Laplacian from the coboundary operator $\delta$.

\noindent\textit{\textbf{Definition 2 (Sheaf Laplacian).}}  
The \emph{sheaf Laplacian} associated with $\mathcal{F}$ is $\mathbf{L}_{\mathcal{F}}=\delta^{\top}\delta : C^{0}(\mathcal{G},\mathcal{F})\rightarrow C^{0}(\mathcal{G},\mathcal{F})\,$,
with block structure
\[
(\mathbf{L}_{\mathcal{F}})_{uu}=\!\!\sum_{e:u\trianglelefteq e}\mathcal{F}_{u\trianglelefteq e}^{\top}\mathcal{F}_{u\trianglelefteq e}\,,
\quad
(\mathbf{L}_{\mathcal{F}})_{uv}=\! -\mathcal{F}_{u\trianglelefteq (u,v)}^{\top}\mathcal{F}_{v\trianglelefteq (u,v)}\,.
\]
The sheaf Laplacian couples the combinatorial structure of $\mathcal{G}$ with the algebraic information encoded by the restriction maps.

\noindent\textit{\textbf{Definition 3 (Local section).}}  
Let $\mathcal{F}$ be a network sheaf over $\mathcal{G}$.  
For an edge $e=(u,v)$, a \emph{local section} is a pair $(x_u,x_v)$ with $x_u\in\mathcal{F}(u)$ and $x_v\in\mathcal{F}(v)$ such that
\[
\mathcal{F}_{u\trianglelefteq e}(x_u)=\mathcal{F}_{v\trianglelefteq e}(x_v)\in\mathcal{F}(e)\,.
\]
Local sections formalize semantic compatibility between agents as perfect agreement of their projections into the communication space. Since global consistency is obtained by enforcing local consistency over edges, semantic alignment over the entire network reduces to enforcing local sections along the selected communication edges. Specifically, learning the sheaf structure from observed latent representations relies on a smoothness prior: agents whose representations are semantically compatible should not vary abruptly across communication edges.  
This leads to minimizing the sheaf total variation $\mathrm{tr}(\mathbf{X}^{\top}\mathbf{L}_{\mathcal{F}}\mathbf{X})$.  
Expanding the trace gives
\[
\Tr{\mathbf{X}^{\top}\mathbf{L}_{\mathcal{F}}\mathbf{X}}
=
\sum_{e=(u,v)\in\mathcal{E}}
\frob{
\mathcal{F}_{u\trianglelefteq e}\mathbf{X}_u
-
\mathcal{F}_{v\trianglelefteq e}\mathbf{X}_v}^{2}\,,
\]
showing that global smoothness corresponds to minimizing semantic inconsistencies across edges.
Constraining the restriction maps to the orthogonal manifold makes the network sheaf equivalent to an unweighted connection graph, and local semantic inconsistencies are measured by $\frob{\myO_{uv}\X_u - \X_v}^2$ \cite{di2024learning}, where $\myO_{uv}$ is an orthogonal matrix for each $e=(u,v)\in\mathcal{E}$.

Although all latent spaces here have the same dimensionality $d$, the effective semantic representations may lie in a lower-dimensional structure, which can be exploited for communication efficiency.  
We therefore introduce a semantic denoising step via a common learnable dictionary $\D \in \reall^{d \times d}$ and directly align the denoised latent representations $\D \eS_i$, where $\eS_i \in \reall^{d \times n}$ are local sparse codes for agent $i \in [V]$.  
The dictionary learning procedure that produces $\D$ and $\eS_i$ is detailed in \Cref{subsec:semantic_denoising}.  
Assuming these denoised representations are given, we can formulate joint learning of the communication topology and restriction maps as a best-subset selection problem.

Given a set of candidate edges, we seek a subset of cardinality $E_{0}$ and orthogonality-constrained restriction maps of the form $\mathcal{F}_{u\trianglelefteq e} = \myO_{uv} \in \orthogonal{d}$ for each edge $e=(u,v)\in\mathcal{E}$, so as to minimize the cumulative semantic inconsistency:
\begin{equation}\label{eq:netsh-learning-prob}
    \begin{aligned}
        \min_{\substack{\{\myO_{uv} \in \orthogonal{d}\},\\ \mya \in \{0,1\}^{V(V-1)/2}}} \quad & \sum_{e \in \edgeset} a_e \frob{\myO_{uv}\D\eS_u - \D\eS_v}^2\,;\\
        \text{subject to} \quad & \norm{\mya}_0 = E_0\,. 
    \end{aligned}
    \tag{P1}
\end{equation}
As shown in \cite{di2024learning}, solving \eqref{eq:netsh-learning-prob} reduces to computing $\myO_{uv}$ for each edge via a local orthogonal Procrustes problem and then selecting the $E_0$ edges with the lowest local loss, corresponding in our case to the highest semantic alignment.

\subsection{Semantic Denoising via Dictionary Learning}\label{subsec:semantic_denoising}

To filter semantic noise and enable information compression, we build a shared semantic space and project the agents' input embeddings onto it before learning network sheaf via \eqref{eq:netsh-learning-prob}. We seek a semantic space that denoises local semantics at the agent level and disentangles their representations. Specifically, our goal is to jointly learn \emph{(i)} a shared semantic space $\D \in \reall^{d \times d}$, and \emph{(ii)} local sparse representations $\eS_i \in \reall^{d \times n}$, $i \in [V]$, obtained as projections of the agents' embeddings onto $\D$. To enhance interpretability, we enforce linear independence among the atoms (columns) of $\D$ through a log-determinant penalty on its Gramian, and constrain each atom $[\D]_k$ to have unit norm for all $k \in [d]$, so that $\D \in \oblique{d}{d}$. To promote parsimony, we impose group sparsity on the rows of $\eS_i$ via the $(2,0)$-norm, effectively selecting $d^\prime_i \leq d$ atoms per embedding. 
The latter $d^\prime_i$ are hyperparameters that can be set in a data-driven manner, e.g., via cross-validation.
Then, letting $\X \coloneqq [\X_1, \ldots, \X_V] \in \reall^{d \times nV}$, $\eS \coloneqq [\eS_1, \ldots, \eS_V] \in \reall^{d \times nV}$, we consider the dictionary learning problem:
\begin{equation}\label{eq:main-dict-problem}
    \begin{aligned}
        \min_{\substack{\D \in \oblique{d}{d}, \eS \in \reall^{d \times nV}}} \; & \frac{1}{2}\frob{\X \!-\! \D\eS}^2 \!-\! \gamma \log\det(\D^\top \D)\,;\\
        \text{subject to} \; & \norm{\eS_i^\top}_{2,0} \leq d^\prime_i, \quad \forall\, i \in [V]\,,
    \end{aligned}
    \tag{P2}
\end{equation}
where $\gamma \in \reall_+$. Problem \eqref{eq:main-dict-problem} is nonconvex and includes nonsmooth terms.  First, $\oblique{d}{d}$ induces a nonconvex constraint on $\D$.  
We handle this noncomvexity using the \emph{splitting of orthogonality constraints} method \cite{lai2014splitting}, relaxing $\D$ to the Euclidean ambient space $\reall^{d \times d}$ and introducing a splitting variable $\myP \in \oblique{d}{d}$ with constraint $\myP- \D = \zeros_{d \times d}$. Second, the objective $\ell(\D, \eS)$ is nonconvex due to the bilinear term in $\frob{\X - \D\eS}^2$ and the log-determinant penalty. To handle this part, we hinge on \emph{successive convex approximation} (SCA) \cite{nedic2018parallel}. Specifically, given the iterate $\M^q \coloneqq \left( \D^q,\eS^q \right)$, a strongly convex surrogate of $\ell$ around $\M^q$ is
\begin{equation}\label{eq:strongly-convex-surrogate}
    \begin{aligned}
        \widetilde{\ell}\left(\D, \eS; \M^q \right)
        &\coloneqq \frac{1}{2}\frob{\X - \D\eS^q}^2 + \frac{1}{2}\frob{\X - (\D^q\eS)}^2 \\
        &\quad + \gamma \Tr{\nabla_{\D}\log\det(\D^\top \D)\at{\M^q}\left(\D - \D^q\right)} \,,
    \end{aligned}
\end{equation}
which satisfies the stationarity conditions $\nabla_{\D}\ell\at{\M^q}=\nabla_{\D}\widetilde{\ell}\at{\M^q}$ and $\nabla_{\eS}\ell\at{\M^q}=\nabla_{\eS}\widetilde{\ell}\at{\M^q}$. Third, to handle the nonsmooth, nonconvex $\ell_{2,0}$-norm constraints in \eqref{eq:main-dict-problem}, we introduce $V$ splitting variables $\Z_i \in \reall^{d \times n}$ with constraints $\Z_i - \eS_i^\top = \zeros_{n \times d}$.  
Indeed, letting $\mathcal{C}_i\coloneqq\{\Z \in \reall^{d \times n} \mid \norm{\Z}_{2,0}\leq d^\prime_i\}$, it is well known that
\begin{equation}\label{eq:proximity-20}
    \prox_{I_{\mathcal{C}_i}}(\Y) \coloneqq \argmin_{\Z \in \mathcal{C}_i} \; \frob{\Z - \Y}^2
\end{equation}
is the matrix $\Z$ retaining the $d^\prime_i$ columns of $\Y$ with largest $\ell_2$-norm. Thus, at each SCA iteration $q$ we solve
\begin{equation}\label{eq:surrogate-problem}
    \begin{aligned}
        \min_{\substack{\D \in \reall^{d \times d}, \myP \in \oblique{d}{d}, \\\eS \in \reall^{d \times nV},\{\Z_i\in \mathcal{C}_i\}}} \quad & \widetilde{\ell}(\D, \eS; \M^q)\,;\\
        \text{subject to} \quad & \myP - \D = \zeros_{d \times d}\,,\\
                                & \Z_i - \eS_i^\top = \zeros_{n \times d}\,, \ \forall\, i \in [V]\,.
    \end{aligned}
    \tag{S1}
\end{equation}  
We now proceed introducing an iterative algorithm for the solution of \eqref{eq:surrogate-problem}. Introducing $\rho \in \reall_+$ and denoting by $\R \in \reall^{d \times d}$ and $\U_i \in \reall^{n \times d}$, $i \in [V]$, the scaled dual variables, the scaled augmented Lagrangian of \eqref{eq:surrogate-problem} is
\begin{equation}\label{eq:aug-L}
    \begin{aligned}
        &\mathcal{L}_\rho(\D, \eS, \myP, \{\Z_i\}) \coloneqq \widetilde{\ell}(\D, \eS; \M^q) + \\
        &+\frac{\rho}{2}\frob{\D - \myP + \R}^2 + \frac{\rho}{2}\sum_i \frob{\eS_i^\top - \Z_i + \U_i}^2\,.
    \end{aligned}
\end{equation}
We then minimize \eqref{eq:aug-L} with respect to the primal variables and maximize with respect to the dual variables via the \emph{alternating direction method of multipliers} (ADMM) \cite{boyd2011distributed}, while enforcing the $\ell_{2,0}$-norm constraints using \eqref{eq:proximity-20}. This yields the following (inexact) SCA updates:
\begin{equation}\label{eq:SCA-update-recursion}
    \begin{aligned}
        \widetilde{\D} &=\left(\X\eS^{q^\top}+\rho(\myP^q - \R^q)+ \gamma \D^q(\D^{q^\top}\D^q)^{-1}\right)\A^{-1}\,, \\
        \D^{q+1} &= \D^q + \alpha^q \left(\widetilde{\D} - \D^q\right)\,, \quad \text{(SCA smoothing step)}\\
        \widetilde{\eS}_i &= \B^{-1}\left(\D^{q^\top}\X_i + \rho\left(\Z_i^{q^\top}-\U_i^{q^\top}\right) \right)\,,\\
        \eS_i^{q+1} &= \eS_i^q + \alpha^q \left(\widetilde{\eS}_i - \eS_i^q\right)\,, \quad \text{(SCA smoothing step)}\\
        [\myP]_k^{q+1} &= \frac{[\D]_k^{q+1} + [\R]_k^q}{\norm{[\D]_k^{q+1} + [\R]_k^q}}\,, \; \forall k \in [d]\,,\\
        \Z_i^{q+1} &= \prox_{I_{\mathcal{C}_i}}\left(\eS_i^{\top^{q+1}} + \U_i^{q+1}\right)\,,\\
        \R^{q+1} &= \R^q + \D^{q+1} - \myP^{q+1}\,,\\
        \U_i^{q+1} &= \U_i^q + \eS_i^{{q+1}^\top} - \Z_i^{q+1}\,;
    \end{aligned}
    \tag{R1}
\end{equation}
where  
(i) $\A \coloneqq \eS^q\eS^{q^\top} + \rho\identity_d$,  
(ii) $\B \coloneqq \D^{q^\top}\D^q + \rho\identity_d$, and  
(iii) $\alpha^q \in \reall_+$ is a diminishing stepsize satisfying the classical stochastic approximation conditions \cite{nedic2018parallel}.
In practice, convergence of \eqref{eq:SCA-update-recursion} is monitored via standard primal and dual residual feasibility conditions \cite{boyd2011distributed}.

\section{Results}\label{sec:results}

In this section, we evaluate our framework in terms of its accuracy–compression trade-off, as well as the interpretability of the learned representations and communication topologies. We consider a network of $V=10$ AI agents, each corresponding to a pretrained image-classification model from the \texttt{timm}\footnote{\href{https://github.com/rwightman/pytorch-image-models.git}{https://github.com/rwightman/pytorch-image-models.git}} library. All agents process the CIFAR-10 dataset (\num{60000} RGB samples, $32\times 32$) and extract $d$-dimensional latent embeddings, with $d = 384$, from the final layer preceding the classification head. The models listed in \Cref{tab:agent_models} span four architectural families, providing a controlled setting to examine semantic heterogeneity and evaluate the methodology in \Cref{sec:methods}\footnote{The source code for the numerical evaluation is available at: \href{https://github.com/SPAICOM/semantic-dict-sheaf.git}{https://github.com/SPAICOM/semantic-dict-sheaf.git}}.

\begin{table}[b]
\centering
\vspace{-1em}
\caption{Deployed AI Agents Overview.}
\resizebox{0.8\columnwidth}{!}{
\begin{tabular}{c l c}
\toprule
\textbf{Agent} & \textbf{Model} & \textbf{Params (M)}\\
\midrule\midrule
0 & vit\_small\_patch16\_224 & 22.05 \\
1 & vit\_small\_patch16\_384 & 22.20 \\
2 & vit\_small\_patch32\_224 & 22.88 \\
3 & vit\_small\_patch32\_384 & 22.92 \\
4 & levit\_128               & 9.21  \\
5 & levit\_conv\_128         & 9.21  \\
6 & levit\_192               & 10.95 \\
7 & efficientvit\_m4         & 8.80  \\
8 & volo\_d1\_224            & 26.63 \\
9 & volo\_d1\_384            & 26.78 \\
\bottomrule
\end{tabular}
}
\label{tab:agent_models}
\end{table}

\begin{figure}[t]
    \centering
    \includegraphics[width=1.0\linewidth]{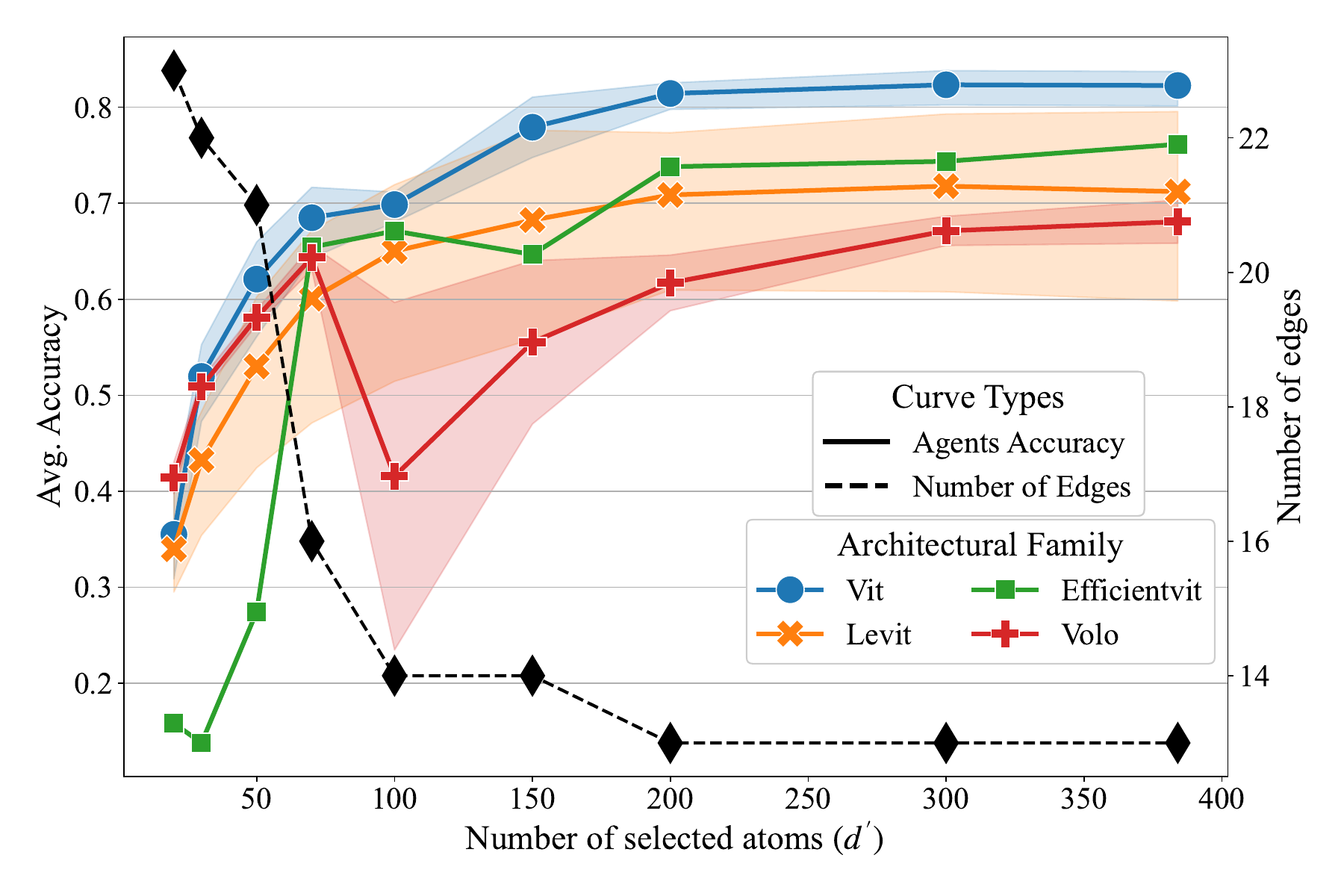}
    \caption{Average semantic accuracy of the architecture families (left y-axis) and communication network edges (right y-axis) along the sparsity level $d'$ for the edge-loss threshold $\tau=0.8$.}
    \label{fig:acc_sparsity}
    \vspace{-1em}
\end{figure}

\subsection{Accuracy-compression trade-off}
The learned semantic space $\D$ enables compression by choosing the sparsity level $d'$ in \eqref{eq:main-dict-problem}.
When agent $v$ communicates with $u$, it transmits the sparse semantic representation $\eS_v$ (with at most $d'$ nonzero rows), obtained by projecting $\X_v$ onto $\D$.
Agent $u$ reconstructs $\X_v$ using $\D$ and interprets it through $\myO_{uv}$.
This allows semantic communication to adapt to channel constraints while mitigating semantic noise.

To measure the effect of compression, we use the \textit{average accuracy} of each agent: the classification accuracy achieved using only the compressed test representations received from its neighbors $\mathcal{N}(v) \subseteq \mathcal{V}$, reconstructed via $\D$ and aligned via $\myO_{uv}$.
\Cref{fig:acc_sparsity} reports this accuracy (left axis) and the number of edges in the pruned topology (right axis) as functions of $d'$, using a fixed edge-loss threshold $\tau=0.8$ on $\frob{\myO_{uv}\D\eS_u - \D\eS_v}^2/\frob{\D\eS_u}^2$. As we can see from Fig. \Cref{fig:acc_sparsity}, accuracy increases with $d'$ across all families, reflecting the reduced compression.
Nevertheless, strong performance is retained even at high sparsity (e.g., $d'=70$).
At the same time, the number of edges grows with sparsity, highlighting that the semantic denoising of $\D$ reduces misalignment losses on both homophilic and heterophilic edges (connecting agents of within and between architectural families, respectively). Hence larger sparsity pushes the representations toward a shared semantic backbone, increasing connectivity; this can also yield local accuracy gains, such as for \textit{Volo} when moving from $d'=100$ to $70$.
Beyond roughly $d'=70$, an oversimplified and increasingly generic semantic dialect emerges. 
This facilitates easier communication between agents, but the expressivity of the language deteriorates rapidly as sparsity continues to increase, ultimately leading to degraded performance in the downstream classification task.

\begin{figure}[t]
    \centering
    \includegraphics[width=1.10\columnwidth, trim=0bp 30bp 0bp 40bp, clip]{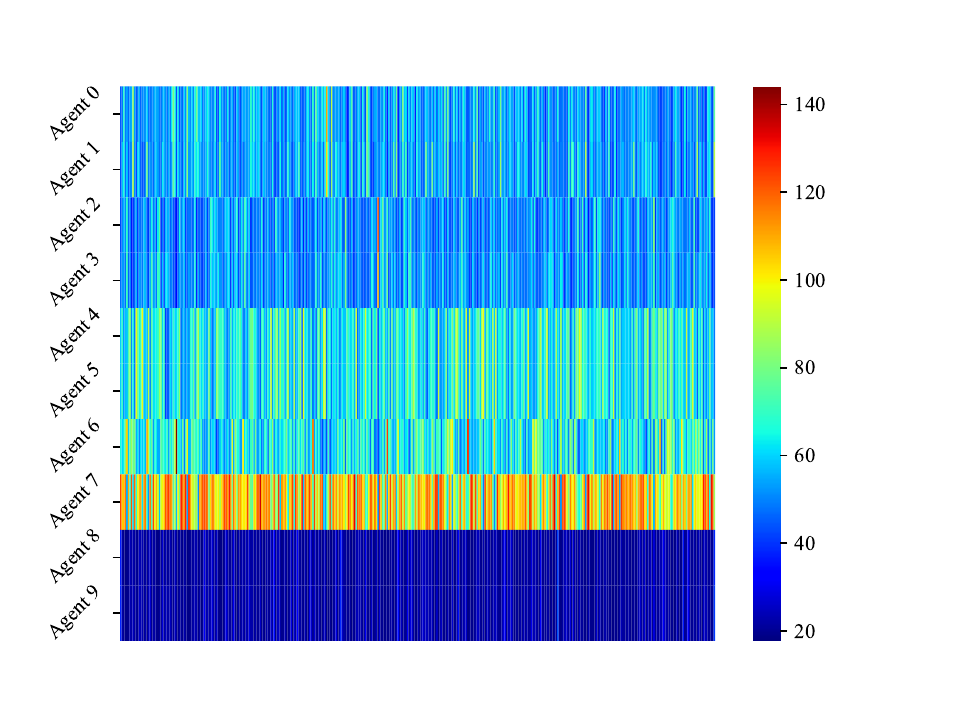}
    \caption{Semantic signatures induced by the global dictionary when setting $d'=d=384$ (no compression), for each model $i=0,\,1,\, \dots,\, 9$ listed in Table~\ref{tab:agent_models}.}
    \label{fig:signatures}
    \vspace{-1em}
\end{figure}

\begin{figure*}[t]
    \centering    \includegraphics[width=.80\textwidth, trim=5bp 150bp 420bp 160bp, clip]{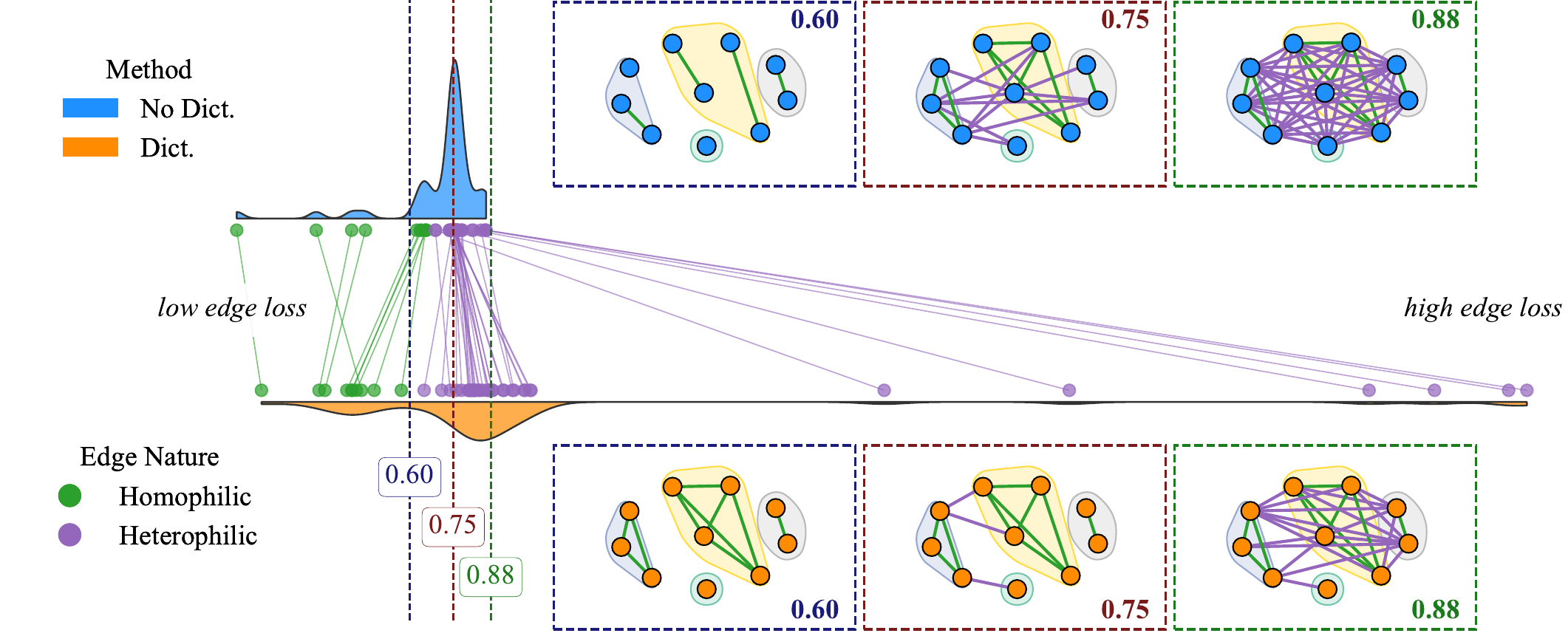}
    \caption{Distribution of edge misalignment losses with (lower panel) and without (upper panel) dictionary learning (shown for $d'=200$), illustrating the emergence of bimodality and improved separation between homophilic and heterophilic edges.}
    \label{fig::edge_distribution}
    \vspace{-1em}
\end{figure*}

\subsection{Semantic denoising for topology learning}

The dictionary $\D$ also enhances interpretability by producing comparable sparse codes across agents.
Even in the non-compressive regime ($d'=d$), the row norms of $\eS_i$ form characteristic \textit{semantic signatures}, see \Cref{fig:signatures}. These signatures quantify how strongly each agent relies on individual dictionary atoms and reveal structural similarities between the agents' latent spaces. For example, \Cref{fig:signatures} shows how \textit{Volo} and \textit{EfficientViT} models' signatures are markedly different from those of all other families, anticipating their role as semantic outliers in the subsequent network sheaf learning stage.

\Cref{fig::edge_distribution} shows that for $d' = 200$, learning $\D$ shifts the edge-loss distribution from unimodal to distinctly bimodal, with heterophilic edges exhibiting higher, more dispersed losses and homophilic edges contracting toward lower values. Consequently, for any fixed loss threshold, the model equipped with $\D$ more effectively suppresses heterogeneous connections while preserving semantically coherent ones, yielding topologies with substantially improved structure. For example, at a threshold of 0.60, the dictionary-based model reconstructs clusters that closely match the ground-truth architectural families, whereas the baseline discards many homophilic edges and produces a highly fragmented graph. At 0.88, the dictionary again filters high-loss heterophilic edges, while the baseline (due to its compact loss distribution) cannot discriminate degrees of heterophily and thus produces an unstructured, densely connected network. Incorporating $\D$ therefore enables principled navigation of the homophily–heterophily spectrum, from strongly homophilic (within-family) edges to the most semantically distant ones. In summary, the semantic denoising induced by $\D$ unfolds the edge-loss landscape, strengthens discrimination among edge types, and preserves topologies that more faithfully reflect ground-truth architectural families, even under significant sparsification.

\section{Conclusion}\label{sec:conclusion}

This paper introduced the first framework for constructing semantic communication networks among heterogeneous AI agents.
The methodology combines a network-sheaf formulation for jointly learning semantic alignment maps and communication topology with a global dictionary that defines a shared semantic space.
The dictionary enables substantial sparsification of the agents’ representations with limited impact on downstream accuracy, and at the same time enhances topology learning by making edge misalignment losses more separable.
As a result, greedy edge-selection strategies become simpler and less sensitive to the exact thresholding value, while still yielding coherent and semantically meaningful communication structures.
The dictionary also induces \emph{semantic signatures} that provide interpretable characterizations of each agent and reflect architectural relationships.
Overall, the results demonstrate the effectiveness of combining semantic denoising with sheaf-based topology learning, and suggest future directions including semantic clustering, adaptive topology control, and efficient protocols for semantic network formation.

 \bibliographystyle{IEEEtran.bst}
\bibliography{IEEEabrv,bibliography}

\end{document}